\documentclass[]{aa}  

\usepackage{graphicx}
\usepackage{txfonts}

\usepackage{tikz,tikz-3dplot}\usetikzlibrary{calc}
\usepackage{mathtools}
\usepackage{caption}
\usepackage{subcaption}
\usepackage{xcolor}
\usepackage{hyperref}
\hypersetup{colorlinks,breaklinks,
  linkcolor=[rgb]{0.368417, 0.506779, 0.709798},
  citecolor=[rgb]{0.368417, 0.506779, 0.709798},
  urlcolor=[rgb]{0.368417, 0.506779, 0.709798}}
\usepackage{siunitx}
\begin{document}

   \title{Dark Matter reconstruction from stellar orbits in the Galactic Centre}

   %\subtitle{I. Overviewing the $\kappa$-mechanism}

   \author{T. Lechien
   	\inst{1}\fnmsep\thanks{Corresponding authors: T.~Lechien (thibaultlechien+aa@gmail.com) and G.~Hei{\ss}el (\mbox{gernot.heissel@esa.int})}
   	\and
   	G. Hei{\ss}el\inst{1}\fnmsep\footnotemark[1]
   	\and
   	J. Grover\inst{1}
   	\and
   	D. Izzo\inst{1}
   }

   \institute{Advanced Concepts Team, European Space Agency, TEC-SF, ESTEC, Keplerlaan 1, 2201 AZ Noordwijk, The Netherlands
              %\email{wuchterl@amok.ast.univie.ac.at}
         %\and
         %    LESIA, Observatoire de Paris, Université PSL, CNRS, Sorbonne Université, Université de Paris, 5 Place Jules Janssen, 92195 Meudon, France\\
             %\email{c.ptolemy@hipparch.uheaven.space}
             %\thanks{}
             }

   \date{Received X; accepted Y}

% \abstract{}{}{}{}{} 
% 5 {} token are mandatory
 
  \abstract
   {Current constraints on distributed matter in the innermost Galactic Centre (such as a cluster of faint stars and stellar remnants, Dark Matter or a combination thereof) based on the orbital dynamics of the visible stars closest to the central black hole, typically assume simple functional forms for the distributions.}
   {We take instead a general model agnostic approach in which the form of the distribution is not constrained by prior assumptions on the physical composition of the matter. This approach yields unbiased - entirely observation driven - fits for the matter distribution and places constraints on our ability to discriminate between different density profiles (and consequently between physical compositions) of the distributed matter.}
   {We construct a spherical shell model with the flexibility to fit a wide variety of physically reasonable density profiles by modelling the distribution as a series of concentric mass shells. We test this approach in an analysis of mock observations of the star S2.}
   {For a sufficiently large and precise data set, we find that it is possible to discriminate between several physically motivated density profiles. However, for data coming from current and expected next generation observational instruments, the potential for profile distinction will remain limited by the precision of the instruments. Future observations will still be able to constrain the overall enclosed distributed mass within the apocentre of the probing orbit in an unbiased manner. We interpret this in the theoretical context of constraining the secular versus non-secular orbital dynamics.}
  % conclusions heading (optional)
   {Our results show that while stellar data over multiple orbits of currently known stars will eventually yield model agnostic constraints for the overall amount of distributed matter within the probe's apocentre in the innermost Galactic Centre, an unbiased model distinction via determining the radial density profile of the distribution is out of the measurement accuracy of the current and next generation instruments in principle. Constraints on Dark Matter models will therefore remain subject to model assumptions, and will not be able to significantly downsize the zoo of candidate models.}


    \maketitle
%
%-------------------------------------------------------------------
\section{Introduction}\label{S: introduction}

Continuous tracking of stellar orbits in the Galactic Centre over the past few decades revealed the massive compact object Sagittarius A* (Sgr~A*) at their shared focal point \citep{NobelCommittee20, Genzel20}. These and other observations, such as of flare orbits \citep{GRAVITY+18_flares, Wielgus+22, GRAVITY+2023_polarimetry_flares} and of the black hole shadow \citep{EHT22}, are in best agreement with Sgr~A* being a massive black hole of about 4 million solar masses, and rule out many alternatives. Assuming the black hole model, the orbit of the star S2 (also S0-2) further allowed observations of relativistic effects, namely of the gravitational redshift \citep{GRAVITY+18_redshift, Do+19, SaidaEtAl2019} and of the Schwarzschild precession \citep{GRAVITY+20_Schwarzschild_prec}.

While there is currently no visible matter between the innermost known S-stars and the accretion flow of Sgr~A*, observations still allow for the presence of both compact and distributed objects in this region, though with strong constraints. Compact matter could for example be an intermediate mass black hole \citep{Tepp+21, GRAVITY+2023_IMBHs}. Distributed matter could be a cluster of faint stars and stellar remnants \citep{JiangLin85, RubilarEckart2001, Merritt2013}, Dark Matter \citep{GondoloSilk1999, GnedinPrimack2004, Sadeghian+2013_DM_distributions, Brito+2020_superradiance} or a combination thereof. Observations currently allow for a few thousand solar masses of distributed matter within a ball of the radius of the apocentre distance of S2 \citep{GRAVITY+22_mass_distribution}, and this upper bound (or detection threshold) is predicted to decrease significantly in the coming years as the apocentre half of the orbit is being observed with the currently most sensitive instruments \citep{Heissel+22}.

Depending on the nature of the distributed matter, theory predicts it to attain certain density profiles. On their full scale, nuclear stellar clusters have been shown to relax to a Bahcall-Wolf cusp \citep{BahcallWolf1976}, that is, to a radial density power law with exponent $-7/4$. However in the context of the innermost Galactic Centre, nuclear clusters are also often modeled as Plummer distributions \citep{Plummer1911} in order to account for a realistically finite core density. Popular candidates for cold Dark Matter around a massive black hole on the other hand are predicted to attain a power law cusp density distribution with exponents in the range from 0.5 to 2.5 constrained by radio and gamma ray observations, which transitions to a core plateau depending on the rate of particle self-annihilation \citep{GondoloSilk1999, Bertone+2002, GnedinPrimack2004, Sadeghian+2013_DM_distributions, Fields+2014, Shapiro+2016, Chan2018, Balaji+2023_DM_spikes_in_gamma_rays, Zuriaga-Puig+2023_HESS_constraints}. The central ‘spike’ of such a cusp can further be softened by the scattering of the Dark Matter by a cluster of faint stars and stellar remnants due to both dynamical heating and capture by the black hole \citep{Merritt2004, BertoneMerritt2005} or as a consequence of a history of hierarchical mergers of Dark Matter halos containing massive black holes \citep{Merritt+2002}. Other forms of Dark Matter again predict different profiles (see for example \citeauthor{Detweiler1980}~\citeyear{Detweiler1980}, \citeauthor{CardosoYoshida2005}~\citeyear{CardosoYoshida2005}, \citeauthor{Dolan2007}~\citeyear{Dolan2007}, \citeauthor{Witek+2013}~\citeyear{Witek+2013}, \citeauthor{Brito+2020_superradiance}~\citeyear{Brito+2020_superradiance} for scalar fields, including types of fuzzy cold Dark Matter \citep{Cardoso+2022}).

Because of this large variety of Dark Matter candidates and models, not all of which are mutually exclusive -- as already noted a combination of a stellar cluster and Dark Matter is plausible -- it is difficult to obtain general constraints concerning the amount, distribution and nature of Dark Matter which might reside in the Galactic Centre. Results such as the quoted current upper observational bound from stellar orbits of a few thousand solar masses come with a prior assumption of the radial density profile of the distribution, and are thus model dependent. In this way, statements about the possible nature of the Dark Matter rely on an incomplete comparison between a zoo of models.

In this work we present a model agnostic approach based on an approximation of Dark Matter profiles by a series of concentric spherical mass shells. The model parameters are the shell masses which are estimated by fitting to observational data. This idea is also used in the mapping of mass concentrations (mascons) for celestial bodies via the geodesy of orbiting spacecrafts \citep{WernerScheerers1996, IzzoGomez22}. The advantage of the mass shell model is that a fit not only estimates the amount but also the radial profile of the underlying density distribution without any priors beyond, in this case, sphericity. Consequently constraints such as observational upper bounds on total Dark Matter mass can be given model independently, and a single model fit can, in principle, discriminate between various physically reasonable Dark Matter models and compositions. In what follows we will illustrate this through the particular example of S2.

In Sect.~\ref{S: Methods} we present our model for a star orbiting a massive black hole through a Dark Matter distribution, and discuss our model and fitting implementation. In Sect.~\ref{S: Results} we use our model to analyse mock observations of S2, focusing on the degree to which the amount and radial profile of distributed matter are constrained by the data. We conclude with a discussion on the implications of our results for future observational Dark Matter constraints in Sect.~\ref{S: Conclusions}.

%--------------------------------------------------------------------
\section{Methods}\label{S: Methods}

\subsection{Equations of motion}\label{SS: Equations of motion}

We consider the scenario of a star (S2) orbiting a massive black hole (Sgr~A*) through a distribution of Dark Matter. The near-Keplerian orbits of S2 allow us to formulate the problem by the equations of motion
\begin{align}\label{E: Newton 2}
    \ddot{\mathbf r} = -\frac{GM_\bullet}{r^2}\mathbf n + \mathbf a_{\mathrm{1PN}} + \mathbf a_{\mathrm{DM}}
\end{align}
where $\mathbf r$ is the position of S2, $r=|\mathbf r|$, $\mathbf n = \mathbf r/r$, $G$ is the gravitational constant, $M_\bullet$ is the black hole mass and a dot denotes time derivation \citep{Merritt2013, PoissonWill2014, Heissel+22}. The first term describes the extreme mass ratio Kepler problem for S2 and Sgr~A*. $\mathbf a_{\mathrm{1PN}}$ takes into account relativity for the two-body problem to first post-Newtonian order (sufficient for S2 with a pericentre distrance of about $1400$ Schwarzschild radii). For our extreme mass ratio case it is given by
\begin{align}\label{E: 1PN acceleration}
    \mathbf a_{\mathrm{1PN}} = 4\frac{GM_\bullet}{c^2r^2}
        \left(
            \left(\frac{GM_\bullet}{r}-\frac{|\mathbf v|^2}{4}\right)\mathbf n +
            \left(\mathbf n\cdot\mathbf v\right)\mathbf v
        \right),
\end{align}
where $\mathbf v = \dot{\mathbf r}$ and $c$ is the speed of light \citep{Merritt2013, PoissonWill2014}. We neglect the very small effects due to black hole spin \citep{Will2008, Zhang+15, Yu+16, Grould+17, Waisberg+18, Qi+21, AlushStone2022, Capuzzo-DolcettaSadun-Bordoni2023}. Here $\mathbf a_{\mathrm{DM}}$ denotes the acceleration exerted by the Dark Matter, which we assume to be sufficiently small for a Newtonian description of $\mathbf a_{\mathrm{DM}}$ to be adequate \citep[Appendix~B]{Heissel+22}. Furthermore we assume the Dark Matter distribution to be spherically symmetric and centered at Sgr~A*.

\subsection{Shell model }\label{SS: Dark Matter shells}

In order to give our Dark Matter model the flexibility to attain a wide variety of physically reasonable radial profiles, we construct it as a sum of concentric shells of radii $(r_i)_{i=1}^N$ and masses $(m_i)_{i=1}^N$. The total mass of the distribution enclosed within a ball of a certain radius $M_C(r)$ then increases in steps with $r$ such that by Newton’s shell theorem
\begin{align}\label{E: DM acceleration}
    \mathbf a_{\mathrm{DM}}(r) =
        - \frac{GM_C(r)}{r^2}\,\mathbf n
\end{align}
with
\begin{align}\label{E: enclosed mass function}
    M_C(r) = \sum_{i=1}^N m_i\frac{1+f(r-r_i)}{2}.
\end{align}
Thin (radially Dirac density distributed) shells correspond to equating $f$ to the Heaviside step function. For differentiability and a better adaptability to physically reasonable profiles, we however smooth out the steps to sigmoids given by the logistic function $f(r-r_i) = \tanh((r-r_i)/\kappa)$, where $\kappa$ controls the smoothness of the increments. Calculating $M_C'(r) = \sum_{i=1}^N (2\kappa)^{-1}m_i\,\mathrm{sech}((r-r_0)/\kappa)^2$ it follows that each Dark Matter shell has a radial density distribution resembling a bell shape of width $\kappa$, marking a roughly $65\%$ drop off. A closely related alternative would be to use the error function for $f$, in which case the Dark Matter shells would have Gaussian radial density distributions.

\subsection{Flexibility of the shell model}\label{SS: Flexibility}

Let us denote the enclosed mass function of a given ground truth (GT) distribution by $M_C^{\mathrm{GT}}(r)$. Then choosing an equal spacing $\Delta r$ between the $r_i$, we define the shell approximation of the ground truth distribution by~\eqref{E: enclosed mass function} with $m_1=M_C^{\mathrm{GT}}(r_1+\Delta r/2)$ and $m_{i+1} = M_C^{\mathrm{GT}}(r_{i+1}+\Delta r/2) - m_i$, similar to a midpoint Riemann sum.

In order to demonstrate the flexibility of our model we consider the three examples
\begin{align}\label{E: ground truth profiles}
    M_C^\mathrm{GT}(r)\propto\begin{cases}
        \bigg(\dfrac{r}{r_0}\bigg)^{5/4} & \text{Bahcall-Wolf cusp} \\
        \bigg(\dfrac{r}{r_0}\bigg)^3\Bigg(1+\dfrac{r^2}{r_0^2}\Bigg)^{-3/2} & \text{Plummer model} \\
        \bigg(\dfrac{r}{\tilde r_0}\bigg)^{11}\Bigg(1 + \dfrac{r^{10}}{\tilde r_0^{10}}\Bigg)^{-11/10} & \text{Zhao model, $\alpha=\tfrac{1}{10}$}
    \end{cases}
\end{align}
where $r_0=2483.1\,\mathrm{AU}$ ($0.3''$ on sky) and $\tilde r_0 = 1200\,\mathrm{AU}$ ($0.15''$ on sky). All profiles are normalised such that $M_C^\mathrm{GT}(r_\mathrm{a}) = 10^{-3}M_\bullet$ where $r_{\mathrm a}$ is the apocentre distance of S2, which corresponds to the current observational upper bound of \citet{GRAVITY+22_mass_distribution}. The distribution of \citet{BahcallWolf1976} resembles a dynamically relaxed nuclear star cluster, but it also serves as a cold Dark Matter spike example (without particle self-annihilation) since its power law exponent lies well within the observational constraints for the respective models, which are as well power laws \citep{GondoloSilk1999, GnedinPrimack2004, Fields+2014, Sadeghian+2013_DM_distributions}. The profile by \citet{Plummer1911} represents a relaxed stellar cluster, which is however also often used to model the innermost region of a nuclear clusters in order to account for a realistically finite central density. The Zhao family of double power law distributions \citet[Table~1]{Zhao1995} provides a parametrization that is flexible enough to capture a variety of analytical DM halo models, and for certain parameter ranges has shown excellent agreement with DM simulations. We  will restrict our attention to a particular instance of this family, the so-called  $\alpha$-model with $\alpha = 1/10$, as this leads to a profile that differs qualitatively from that of Plummer and Bahcall-Wolf.

\begin{figure*}%[hbt]
\centering
\begin{subfigure}{0.33\textwidth}
  \centering
  \includegraphics[width=\textwidth]
  {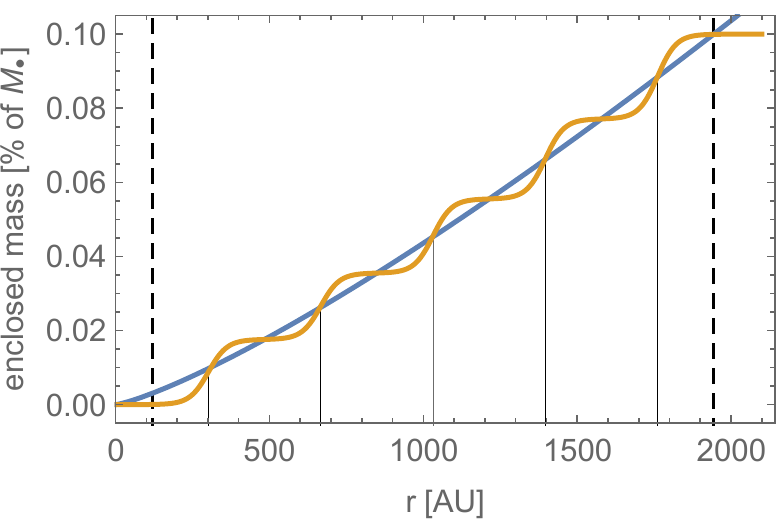}%[clip=true,trim=0 0 0 0,scale=.55]
  \subcaption{Bahcall-Wolf cusp approx. with $\kappa=6$.}\label{F: encl_mass_BW_5_6}
\end{subfigure}
\begin{subfigure}{0.33\textwidth}
  \centering
  \includegraphics[width=\textwidth]
  {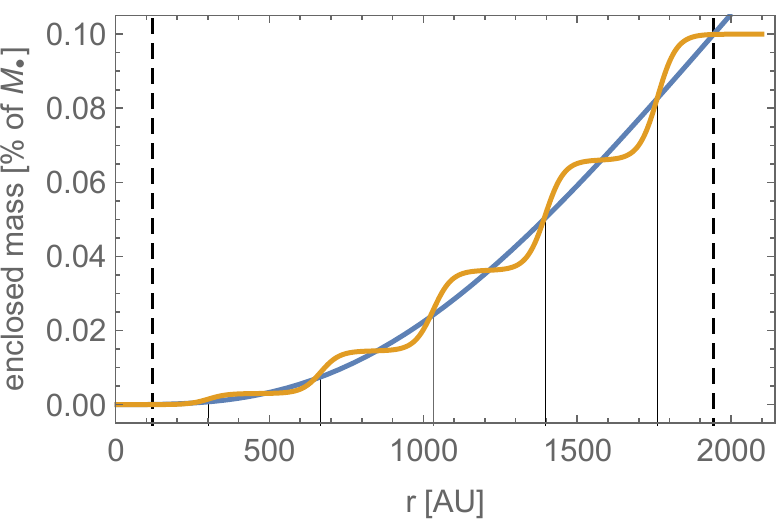}%[clip=true,trim=0 0 0 0,scale=.55]
  \subcaption{Plummer profile approx. with $\kappa=6$.}\label{F: encl_mass_Plm_5_6}
\end{subfigure}
\begin{subfigure}{0.33\textwidth}
  \centering
  \includegraphics[width=\textwidth]
  {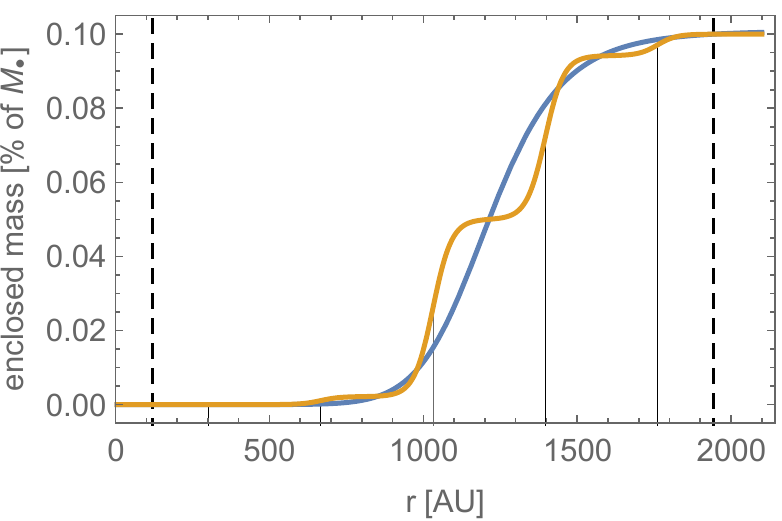}%[clip=true,trim=0 0 0 0,scale=.55]
  \subcaption{Zhao profile approx. with $\kappa=6$.}\label{F: encl_mass_alpha_5_6}
\end{subfigure}
\\
\begin{subfigure}{0.33\textwidth}
  \centering
  \includegraphics[width=\textwidth]
  {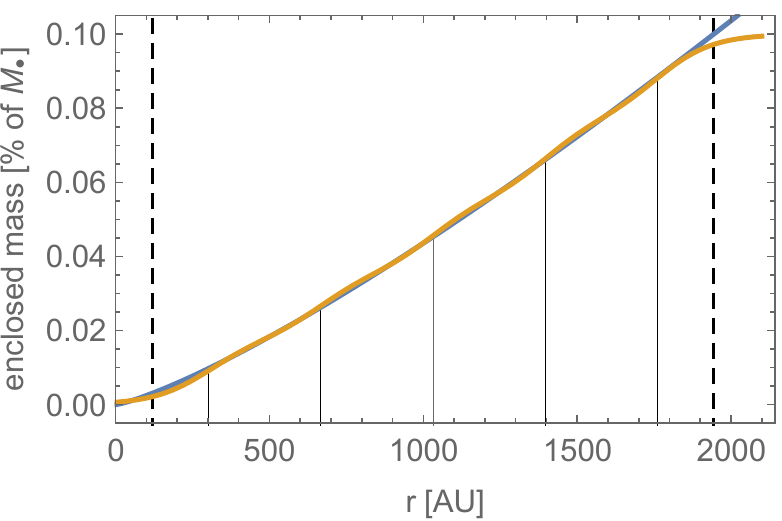}%[clip=true,trim=0 0 0 0,scale=.55]
  \subcaption{Bahcall-Wolf cusp approx. with $\kappa=2$.}\label{F: encl_mass_BW_5_2}
\end{subfigure}
\begin{subfigure}{0.33\textwidth}
  \centering
  \includegraphics[width=\textwidth]
  {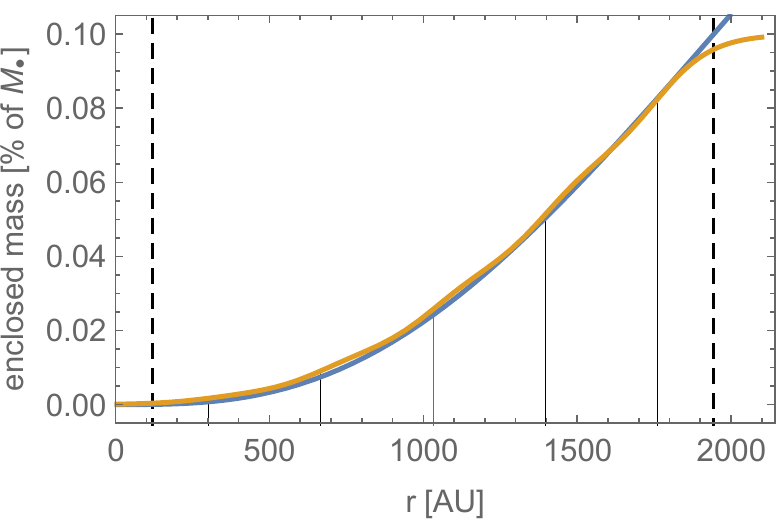}%[clip=true,trim=0 0 0 0,scale=.55]
  \subcaption{Plummer profile approx. with $\kappa=2$.}\label{F: encl_mass_Plm_5_2}
\end{subfigure}
\begin{subfigure}{0.33\textwidth}
  \centering
  \includegraphics[width=\textwidth]
  {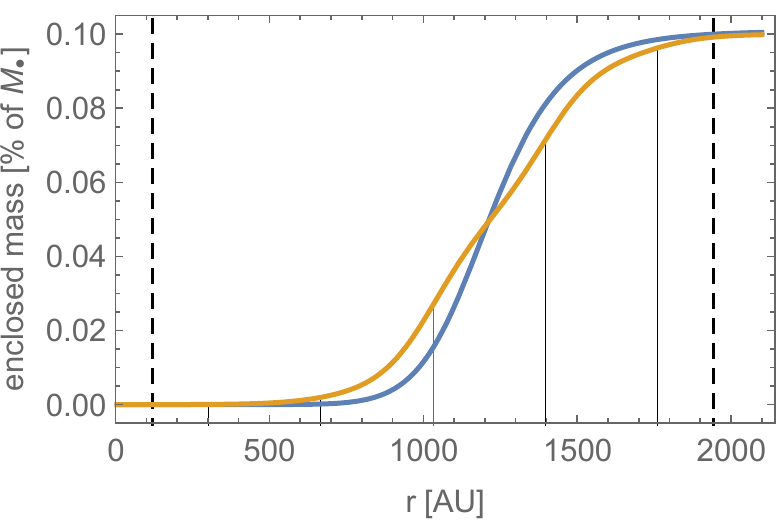}%[clip=true,trim=0 0 0 0,scale=.55]
  \subcaption{Zhao profile approx. with $\kappa=2$.}\label{F: encl_mass_alpha_5_2}
\end{subfigure}
\caption{Approximations with five mass shells (orange) for $\kappa=6$ (top) and $\kappa=2$ (bottom) of three different ground truth profiles (Bahcall-Wolf (left), Plummer (centre), Zhao (right)). Solid vertical lines mark the centre positions of the shells $r_i$. Dashed lines mark the pericentre and apocentre of S2.}
\label{F: shell approximations}
\end{figure*}

Figure~\ref{F: shell approximations} shows two approximations with five mass shells to the above ground truth models.
In the $\kappa=6$ approximation the individual mass shell distributions are clearly separated resulting in well pronounced enclosed mass steps. In the $\kappa=2$ approximation the mass shells have more overlap and the enclosed mass function increases smoothly, resulting in a better approximation of the example profiles---the errors are smaller than $5\times10^{-3}$ (Bahcall-Wolf), $8\times10^{-3}$ (Plummer) and $2\times10^{-2}$ (Zhao) percent of $M_\bullet$.

\subsection{Orbit model}\label{SS: Orbit model}

Equations~\eqref{E: Newton 2} to~\eqref{E: enclosed mass function} together complete our model for the physical stellar orbit. The apparent orbit in terms of the observables right ascension (RA), declination (DEC) and radial velocity (RV), we obtain by projecting the physical orbit onto the coordinate system of the observer (see e.g. \citet[Fig.~1 and Eqs.~(A.7)-(A.8)]{Heissel+22}). The transformation from physical to angular distances thereby introduces the distance of the observer on earth to the Galactic Centre $R_0$ as an additional parameter. For simplicity we neglect observational effects such as the R\o mer delay, the gravitational redshift, the transverse Doppler shift or the motion of the Solar system \citep{Grould+17}. While these are crucial for fitting a model to real observational data, they would only unnecessarily complicate the model for our mock data analysis of Sect.~\ref{S: Results}, in which we focus on whether it is possible to discriminate, in principle in the most straightforward cases, between Dark Matter models.

In summary our model produces the $3$ observables $\mathrm{RA}$, $\mathrm{DEC}$, $\mathrm{RV}$ at given observation times and for given model parameters. Of these parameters we will always fix the (equally spaced)  shell positions $r_i$ and the shell widths $\kappa$. For simplicity we will also fix the black hole mass $M_\bullet$ and the distance to the Galactic Centre $R_0$. This leaves our model with $6 + N$ free parameters to be estimated by fitting to data: $6$ initial conditions for Eq.~\eqref{E: Newton 2} and the $N$ shell masses $m_i$ of Eq.~\eqref{E: enclosed mass function}, which enter the equations of motion (Eq.~\eqref{E: Newton 2}) via the Dar Matter acceleration term (Eq.~\eqref{E: DM acceleration}). For all our case studies of Sect.~\ref{S: Results} we choose $N=5$ and $\kappa=2$ for both the creation of the mock observations as well as for the fitting of the model to these data.

\subsection{Model implementation and fitting algorithm}\label{SS: implementation and fitting}

Instead of integrating equation~\eqref{E: Newton 2} directly, we use the equivalent system of osculating equations in the form of \citet[Eqs.~(3.64)-(3.66)]{PoissonWill2014} to obtain our model orbits. These are a system of six evolution equations in time for the following set of osculating orbital elements: the semi-latus rectum $p$, the eccentricity $e$, the inclination $\iota$, the argument of pericentre $\omega$, the argument of the ascending node $\Omega$ and the true anomaly $f$ of the osculating orbits. The initial elements we denote by a subscript 0. The transformation law from the position and velocity to the elements is given by~\citet[Eqs.~(3.40) to (3.41)]{PoissonWill2014}. For the integration we utilize the open source higher-order Taylor integrator \texttt{heyoka} by \citet{BiscaniIzzo2021}. 
% @Gernot:
%Original text:
%To fit our model to data we use gradient descent of least squares weighted by the measurement uncertainties, and we simultaneously integrate the variational equations, given by the gradient of the model function with respect to the model parameters (i.e. the model sensitivities), in order to assist the algorithm. For this we use the Adam optimiser by \cite{KingmaBa2015} with the recommended default parameters and a learning rate of $10^{-5}$.
% ----
% Alternative explanation:
To fit our model to data we use gradient descent of least squares weighted by the measurement uncertainties. For this we use the Adam optimiser by \cite{KingmaBa2015} with the recommended default parameters and a learning rate of $10^{-5}$. The gradient of the model function with respect to the model parameters (i.e. the model sensitivities) is given by variational equations, which are simultaneously integrated with the model function. 
% Optional add-on (available for just €4.99):
% In each iteration, the model error is calculated and the initial guesses for both the initial conditions and the shell masses are improved in the negative direction of the gradient.

% -------------------------------------------------------------------

\section{Results}\label{S: Results}

\subsection{Ideal data: vanishing measurement uncertainties}\label{SS: Ideal data}

\begin{figure}%[hbt]
  \centering
  \includegraphics[width=\columnwidth]{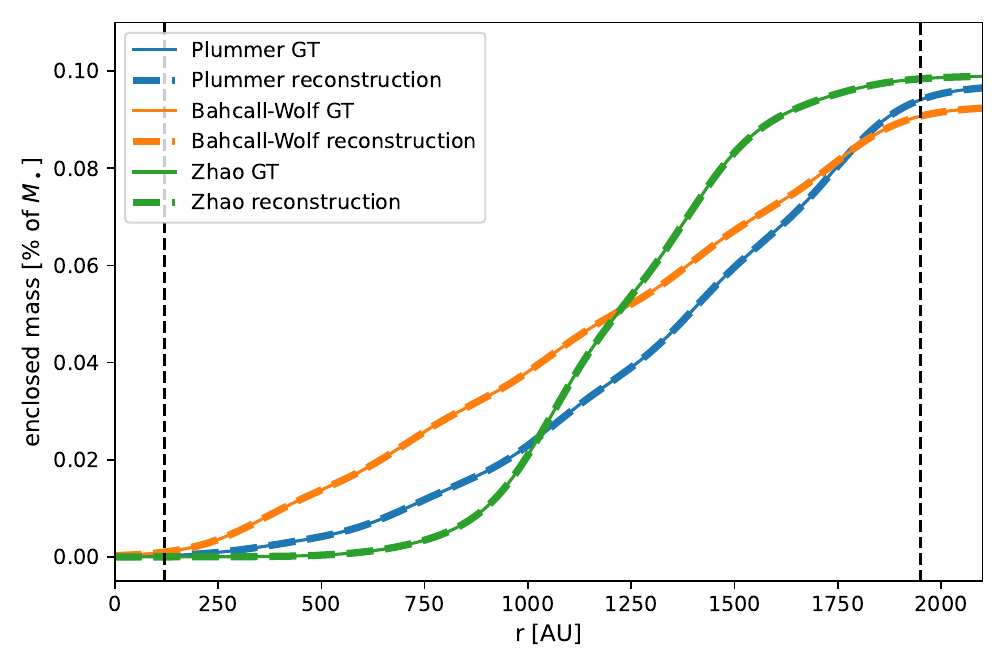}%[clip=true,trim=0 0 0 0,scale=.55],  
  \caption{Reconstructed enclosed mass curves corresponding to the best fits to 300 mock observations with vanishing instrument uncertainties. The three sets of mock data are created with the mass shell approximations (solid lines) of the profiles of Eq.~\eqref{E: ground truth profiles}. The fits are initiated from IG$_1$ (Table~\ref{T: ground truth parameters}).}
\label{F: noiseless}
\end{figure}

For the ground truth which generates our mock observations we take our model of Sect.~\ref{SS: Orbit model} with the (GT) parameter values of Table~\ref{T: ground truth parameters} and the Dark Matter profiles of equation~\eqref{E: ground truth profiles} approximated by five mass shells with $\kappa=2$ (Sect.~\ref{SS: Flexibility} and Figs.~\ref{F: encl_mass_BW_5_2} to~\subref{F: encl_mass_alpha_5_2}).
\begin{table*}[h]
\caption{Ground truth and initial guess.}
\label{T: ground truth parameters}
\centering
\begin{tabular}{l|lllllllll}
 & $M_\bullet$ [$10^6 M_\odot$] & $R_0$ [pc] & $p_0$ [AU] & $e_0$ [1] & $\iota_0$ [°] & $\Omega_0$ [°] & $\omega_0$ [°] & $f_0$ [°] & $m_i$ [$M_\bullet$] \\
\hline
GT & 4.297 & 8277 & 225.271 & 0.88441 & -134.70 & 228.19 & 66.25 & -180 & Sect.~\ref{SS: Flexibility} \\
IG$_1$ & fixed & fixed & GT & GT & GT & GT & GT & GT & 0 \\
IG$_2$ & fixed & fixed & GT $\pm$ 0.0131 & $\mathrm{GT}\pm6\cdot10^{-6}$ & GT $\pm$ 0.003 & GT $\pm$ 0.003 & GT $\pm$ 0.003 & GT & 0 \\
IG$_3$ & fixed & fixed & GT $\pm$ 0.1310 & $\mathrm{GT}\pm6\cdot10^{-5}$ & GT $\pm$ 0.030 & GT $\pm$ 0.030 & GT $\pm$ 0.030 & GT & 0
\end{tabular}
\tablefoot{GT = ground truth. IG = initial guess. Where IG is given as an interval, it is understood as a normal distribution with mean $\pm$ standard deviation from which the initial guess is drawn. For $p_0$, $e_0$, $\iota_0$, $\Omega_0$ and $\omega_0$ we take the observational best fit values of \citet[Table~B.1]{GRAVITY+22_mass_distribution} for GT and the corresponding parameter uncertainties as standard deviations for IG$_3$, and $1/10$ thereof for IG$_2$. Our inclination has the opposite sign to that of our source due to different conventions \citep[Appendix~C]{Heissel+22}. For simplicity we do not fit but fix the values for $M_\bullet$ and $R_0$. Without loss of generality we choose our initial orbit of the GT to osculate at apocentre ($f_0=-180$°). The fact that we do not vary the IG for $f_0$ does not compromise our procedure since its value still varies during the fitting process.}
\end{table*}
We start our investigation with 300 (RA, DEC, RV) mock observations equally time distributed over one orbital period of S2. This is a conservative estimate, taking into account down time for commissioning etc., for the amount of data that will be gathered if the current observing rate with the Very Large Telescope's (VLT) GRAVITY interferometer \citep{GRAVITY+17_first_light} and ERIS spectrometer \citep{Davies+2018_ERIS} continues. We then perform fits as described in Sect.~\ref{SS: implementation and fitting}. In the idealised scenario of vanishing instrument uncertainties and starting from the initial guess IG$_1$ of Table~\ref{T: ground truth parameters} our model is able to reconstruct the three ground truth distributions from the fits up to machine precision (Fig.~\ref{F: noiseless}).
This is a proof of principle, in that we initialise mascon masses to vanish, hence assume no prior information on the underlying distribution, and fit a model flexible enough to have local optima away from the ground truth, yet nevertheless attain it. 

\subsection{Idealised data: measurement uncertainties of 1/10th of the current instrument precision}\label{SS: Idealised data}

Introducing Gaussian noise to the observable (RA, DEC, RV) data to simulate measurement uncertainties, we find that the reconstructed enclosed mass profiles now vary depending both, on the noise realisation of the data (that is, on the specific data instance for the observables drawn from the statistical distribution) and on the initial parameter guess. We also find that a covariance analysis yields small uncertainties in the estimated shell masses. Taken together these points suggest that the problem is not convex such that the least squares map now indeed exhibits multiple local minima to which the algorithm can descend, or narrow valleys in which it stagnates. One likely reason for this are degeneracies between neighbouring mass shells; an overestimated $m_i$ can, within measurement uncertainties, be balanced by an underestimated $m_{i+1}$. To control for this statistically, we perform a sample of 10 fits per ground truth, where with each fit we vary not only the noise realisation (that is, for each fit we draw new (RA, DEC, RV) samples from Gaussian distributions) but also the initial guess for the parameters within reasonable bounds. For these bounds we orient ourselves at the current observational constraints of \citet[Table~B.1]{GRAVITY+22_mass_distribution} and draw the initial orbital elements from the normal distributions IG$_2$ or IG$_3$ of Table~\ref{T: ground truth parameters}, depending on the chosen measurement uncertainties. The shell masses on the other hand we always guess to vanish when initiating the fits. While the factor ten difference between the standard deviations of IG$_2$ and IG$_3$ is chosen ad~hoc, the trend is motivated by the fact that smaller measurement uncertainties would also yield narrower bounds for the parameter estimates.

\begin{figure*}%[hbt]
  \centering
  \begin{subfigure}{\columnwidth}
    \centering
    \includegraphics[width=\columnwidth]{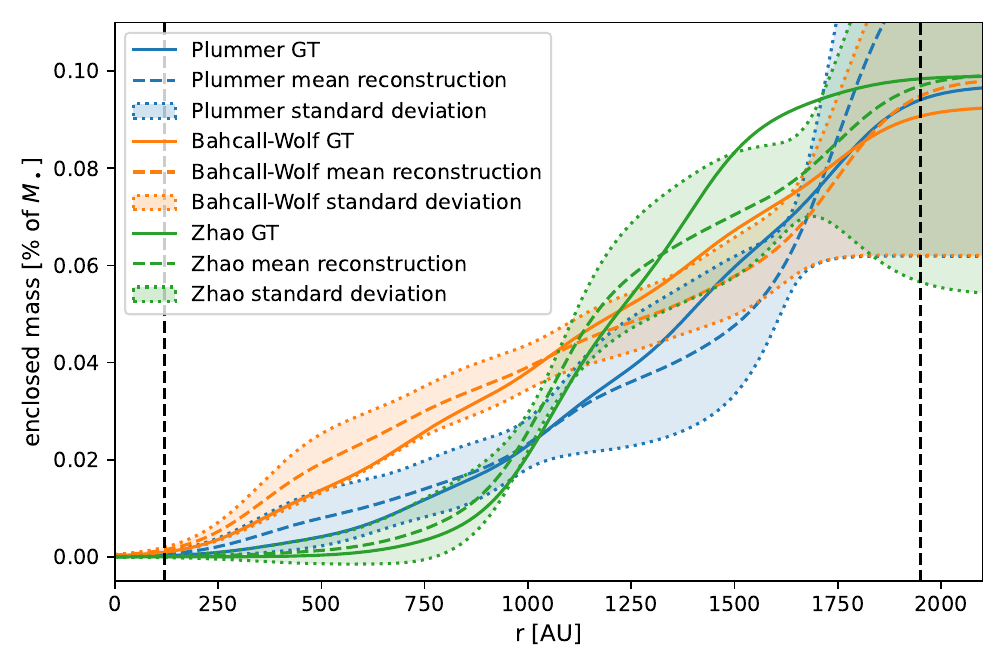}%[clip=true,trim=0 0 0 0,scale=.55],
    \caption{$\mathrm{\Delta RA, \Delta DEC=5\mu''; \Delta RV=1\,km/s; 300\text{ obs. over 1 period}}.$}
    \label{F: 1/10th noise}
  \end{subfigure}
  \begin{subfigure}{\columnwidth}
    \centering
    \includegraphics[width=\columnwidth]{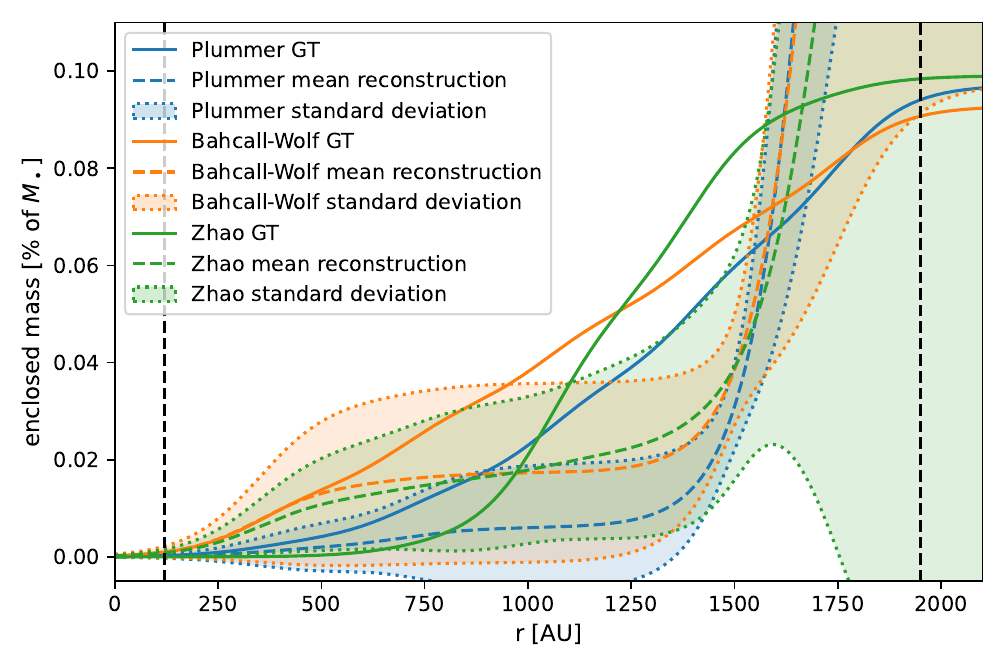}
    \caption{$\mathrm{\Delta RA, \Delta DEC=50\mu''; \Delta RV=10\,km/s; 300\text{ obs. over 1 period}}.$}
    \label{F: realistic noise}
  \end{subfigure}
  \\
  \begin{subfigure}{\columnwidth}
    \centering
    \includegraphics[width=\columnwidth]{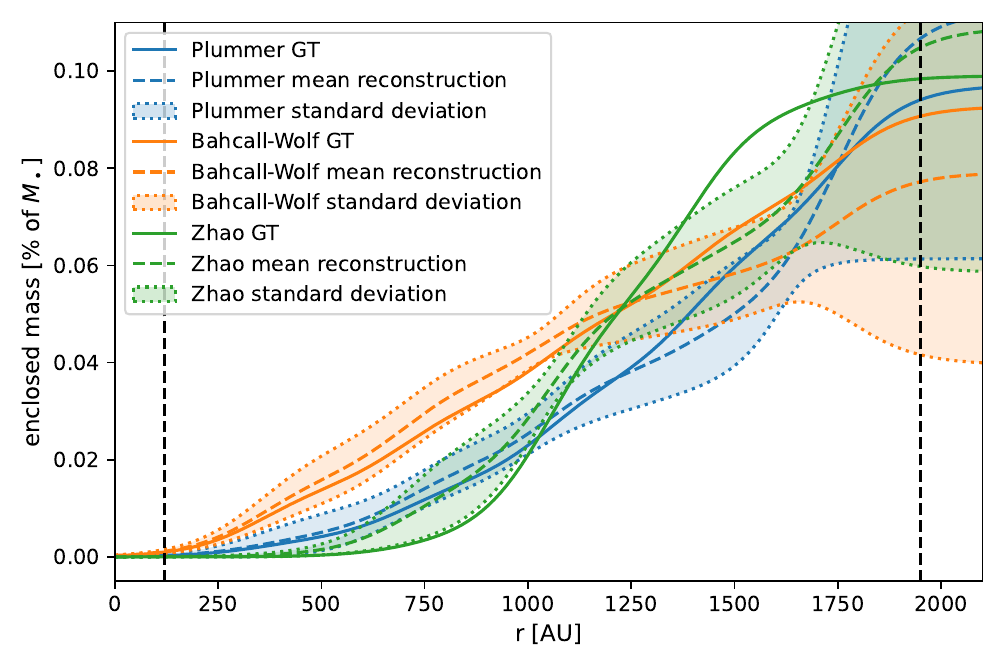}
    \caption{$\mathrm{\Delta RA, \Delta DEC=5\mu''; \Delta RV=1\,km/s; 3000\text{ obs. over 1 period}}.$}
    \label{F: 1/10th noise, 3000 obs, 1 orbit}
  \end{subfigure}
  \begin{subfigure}{\columnwidth}
    \centering
    \includegraphics[width=\columnwidth]{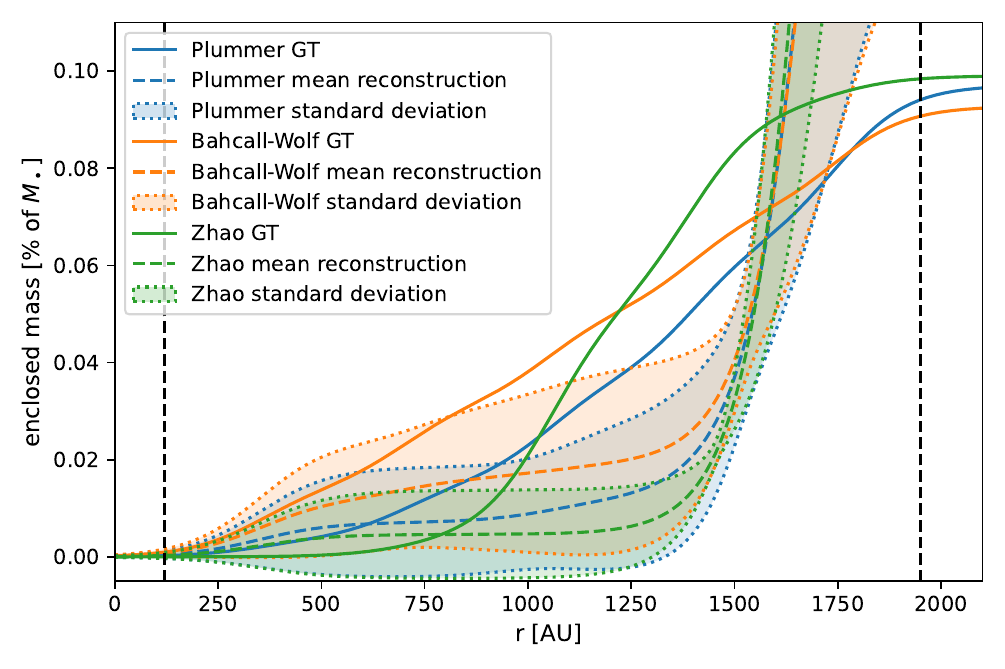}
    \caption{$\mathrm{\Delta RA, \Delta DEC=50\mu''; \Delta RV=10\,km/s; 3000\text{ obs. over 1 period}}.$}
    \label{F: 1 noise, 3000 obs, 1 orbit}
  \end{subfigure}
  \\
  \begin{subfigure}{\columnwidth}
    \centering
    \includegraphics[width=\columnwidth]{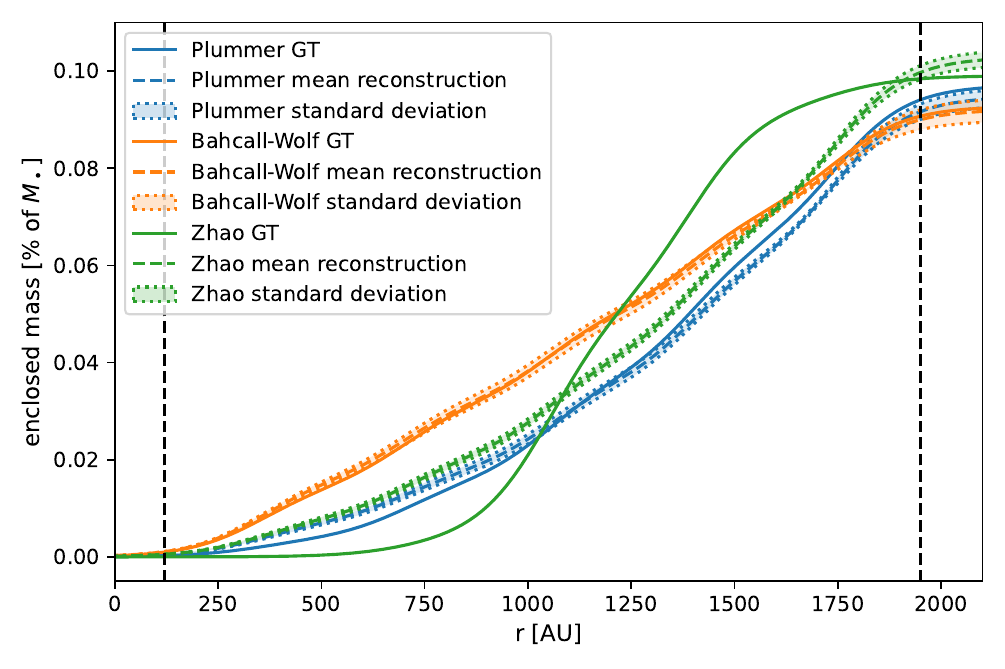}
    \caption{$\mathrm{\Delta RA, \Delta DEC=5\mu''; \Delta RV=1\,km/s; 3000\text{ obs. over 10 period}}.$}
    \label{F: 1/10th noise, 3000 obs, 10 orbits}
  \end{subfigure}
  \begin{subfigure}{\columnwidth}
    \centering
    \includegraphics[width=\columnwidth]{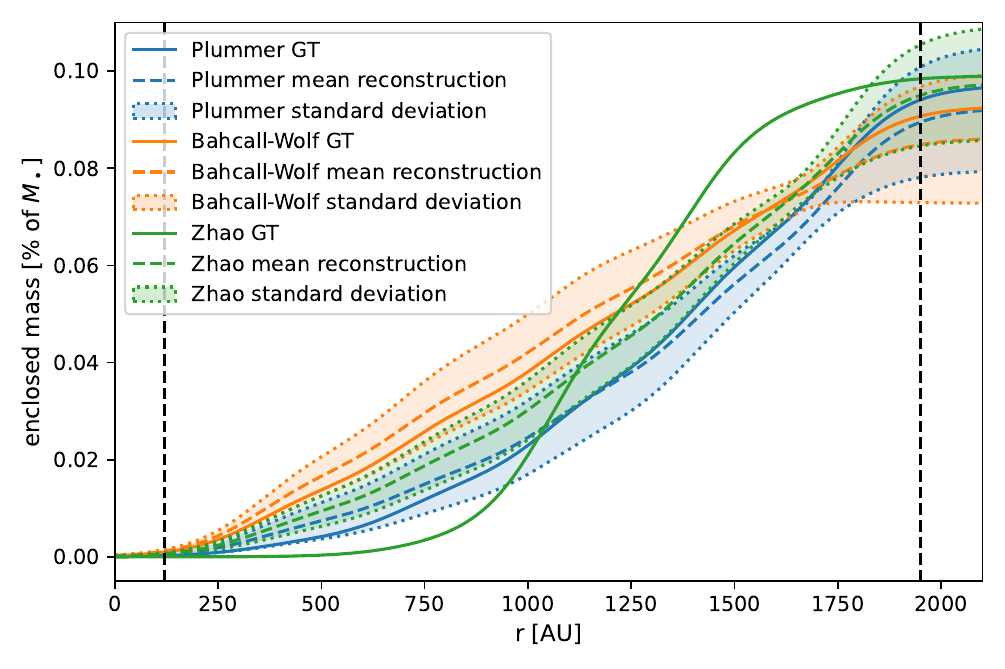}
    \caption{$\mathrm{\Delta RA, \Delta DEC=50\mu''; \Delta RV=10\,km/s; 3000\text{ obs. over 10 period}}.$}
    \label{F: 1 noise, 3000 obs, 10 orbits}
  \end{subfigure}
  \caption{Statistics of reconstructed enclosed mass curves resulting from ten fits per ground truth case (Eq.~\eqref{E: ground truth profiles}) to different mock data. Left: measurement uncertainties of $5\mu''$, $1\,\mathrm{km/s}$ and initial guess drawn from IC$_2$ of Table~\ref{T: ground truth parameters}. Right: measurement uncertainties of $50\mu''$, $10\,\mathrm{km/s}$ and initial guess drawn from IC$_3$ of Table~\ref{T: ground truth parameters} (right). Top: 300 mock observations over one orbital period. Mid: 3000 mock observations over one orbital period. Bottom: 3000 mock observations over ten orbital periods. Means (dashed); standard deviations (shaded); shell approximated ground truth profiles (solid).}
  \label{F: noisy cases}
\end{figure*}

\begin{figure*}%[hbt]
    \centering
    \includegraphics[width=\textwidth]{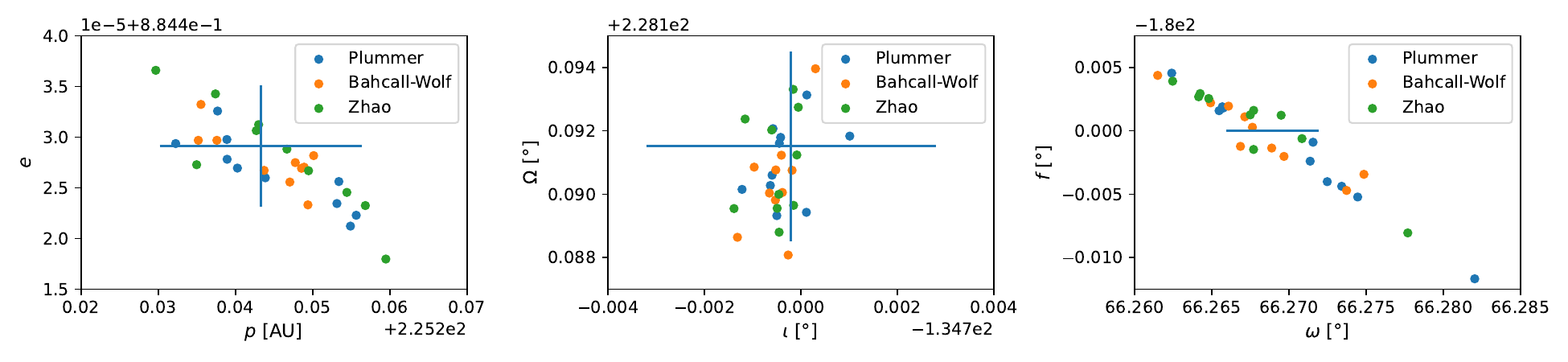}%[clip=true,trim=0 0 0 0,scale=.55],
    \caption{Best fit values (dots) for the initial osculating orbital elements corresponding to the cases of Fig.~\ref{F: 1/10th noise}. The cross in each plot marks the mean and standard deviation of the normal distribution IC$_2$ from which the initial guesses for the fits have been drawn. The corresponding plots for all other considered cases look qualitatively the same.}
    \label{F: orbital elements}
\end{figure*}

Choosing measurement uncertainties of one tenth of the current instrument performances in measuring S2, that is $5\,\mu''$ in astrometry and $1\,\mathrm{km/s}$ in RV \citep{GRAVITY+20_Schwarzschild_prec, GRAVITY+22_mass_distribution}, and drawing the initial guess from IG$_2$ we obtain Fig.~\ref{F: 1/10th noise}.
We see that while even for this low noise level the reconstructed enclosed mass profiles spread, they still cluster around the correct respective ground truths. In particular, for low to mid radii the model is able to clearly distinguish between the three profiles. For large radii the fits are however less constrained, in particular for radii greater than about $1600\,\mathrm{AU}$ -- the domain of the outermost shell. Consequently, the overall amount of distributed matter within the apocentre of S2 is poorly constrained. We interpret this as follows. Shells located at lower radii influence the orbit over longer timescales, and are thus constrained by more data over time than shells located at larger radii. From Fig.~\ref{F: orbital elements} we see that the initial osculating orbital elements converge mostly within the statistics from which they have been drawn initially.
We also observe two correlations. A larger initial eccentricity preferably pairs with a smaller initial semi-latus rectum, which can be understood from the fact that $p\propto(1-e^2)$. That a larger initial true anomaly preferably pairs with a smaller initial argument of pericentre follows directly from the geometric relation between the two angles (see for example \citet[Fig.~1]{Heissel+22}). The respective plots for the orbital elements for all the cases that follow look qualitatively the same as in this case.

Increasing the number of observations to 3000 while keeping them confined within a single orbit and repeating the above procedure results in Fig.~\ref{F: 1/10th noise, 3000 obs, 1 orbit}. We see that the tenfold increase in observations only yields an insignificant narrowing of the spread of the reconstructions in the low to mid radii regime, and no improvement in the large radii regime, including the estimation of the overall amount of distributed mass within apocentre. In contrast, spreading the 3000 observations over ten orbital periods we see a strong improvement in the spread of the reconstructions for all radii (Fig.~\ref{F: 1/10th noise, 3000 obs, 10 orbits}). Both the Bahcall-Wolf and the Plummer GT profiles are reconstructed unambiguously, however the fits failed to reconstruct the Zhao profile. Redoing this case with an initial guess for the $m_i$ close to their GT values, does recover the Zhao profile. We conclude that in the Zhao case, the gradient descent, initiated from $m_i = 0$, gets trapped in a local optimum on the least squares map, with a value around that of the GT optimum. Remarkably however, even though the profile of the Zhao distribution is not found when initiating from $m_i=0$, the estimate on the overall enclosed mass within apocentre is still precise and accurate.

We can understand this when we think of the orbital dynamics in terms of its secular and non-secular components \citep{Merritt2013}. The former are changes in the orbital elements which accumulate per orbit. The latter are changes in the orbital elements which average to zero per orbit. In our present case there is only one secular effect on the orbit caused by the Dark Matter, which is a (negative) pericentre advance $\partial_t\langle\omega\rangle_t$, adding a retrograde component to the precession of the orbit within its plane \citep{JiangLin85, Merritt2013, RubilarEckart2001, Heissel+22}. (A prograde component is cause by the first post-Newtonian correction~\citep{Merritt2013, PoissonWill2014, GRAVITY+20_Schwarzschild_prec}). The non-secular effects caused by the distributed matter correspond to the small variations in the orbital elements over time, and consequently to the small distinct distortions from a Kepler ellipse, which the different profiles inscribe into each orbit. In the case of Figs.~\ref{F: 1/10th noise}, \ref{F: 1/10th noise, 3000 obs, 1 orbit} the measurement uncertainty is sufficient to capture the non-secular effects for low and mid radii, and therefore allows to differ between the signatures which each GT profile inscribes into the single orbit. The lack of data over multiple orbits however prevents good constraints on the pericentre advance per orbit, and therefore on the overall enclosed mass within apocentre. In the case of Fig.~\ref{F: 1/10th noise, 3000 obs, 10 orbits} on the other hand, the data over multiple orbits allows good constraints on both, profile and amount, in two out of three cases. In the Zhao case on the other hand a good fit was found with a local optimum corresponding to a profile distinct from that of the GT. This shows that different profiles and hence different non-secular signatures are compatible with the same data within measurement uncertainties. Since these measurement uncertainties have so far been underestimated, this strongly suggests that an unambiguous constraining of the radial profile of an underlying distribution by future observations with the current and next generation instruments is not possible. What however appears very feasible with future data is to put unbiased constraints on the overall amount of enclosed mass within apocentre, that is, constraints which are not subject to an a~priori assumption on the functional form of the distribution. The fact that the amount of mass is estimated well even in the Zhao case of Fig.~\ref{F: 1 noise, 3000 obs, 10 orbits} lends credence to this conclusion as do the results that follow for realistic measurement uncertainties.

\subsection{Realistic data: measurement uncertainties of the current instrument precision}\label{SS: Realistic data}

The right panel of Fig.~\ref{F: noisy cases} shows the results for the same sequence of cases as in the left panel (Sect.~\ref{SS: Idealised data}), however with realistic measurement uncertainties in the mock data of $50\,\mu''$ and $10\,\mathrm{km/s}$, that is, with the current performances of GRAVITY, SINFONI and ERIS in tracking S2. Clearly, with data confined to one orbit the fits fail to reconstruct any of the ground truth distributions, and there are also no statistical trends, not even for low radii (Figs.~\ref{F: realistic noise}, \subref{F: 1 noise, 3000 obs, 1 orbit}). The measurement uncertainties are simply too large to capture the non-secular variations within a single orbit. For 3000 observations over 10 orbits clear trends again emerge, such that the Plummer and Bahcal-Wolf profiles can be reconstructed unambiguously, though with a significantly larger spread as in the case of Fig.~\ref{F: 1/10th noise, 3000 obs, 10 orbits}. The fits, however, still fail to reconstruct the profile of the Zhao GT and find another locally optimal profile instead. The enclosed mass within apocentre is again well constrained for all cases, including the Zhao model, given the increased measurement uncertainty. This indicates that future observations over several orbits will indeed allow for unbiased constraints on the overall amount of enclosed mass within apocentre.

% -------------------------------------------------------------------

%--------------------------------------------------------------------
\section{Conclusions}\label{S: Conclusions}

\subsection{Summary}

In this work we presented a novel approach to model spherically symmetric matter distributions by concentric mass shells. We implemented this concept into a relativistic model for a star (S2) orbiting a massive black hole (Sgr~A*) through a Dark Matter distribution (cluster of faint stars and stellar remnants, Dark Matter or a combination thereof). We then performed a proof of principle that, given astrometric and spectroscopic observations of sufficient precision, a fit of our model is able to find an underlying ground truth Dark Matter distribution with high enough accuracy to discriminate between various physical models and compositions. Unlike with conventional models based on certain functional forms for the density distribution, no a priori assumption about the principal profile of the ground truth is made.

% @Gernot:
% The first sentence of the paragraph below seems kind of strong, in the sense that we showed it is not possible with 1 star, with our model. I don't think this rules out that it is possible with more stars and a different model?
% In the second sentence, I would change 'will yield' to 'can yield'
Using the spherical shell model our results indicate that with the measurement uncertainty of current and next generation instruments it will not be possible to achieve unbiased constraints on the radial profile of an underlying matter distribution. We however showed that future observations can yield unbiased constraints on the overall amount of enclosed mass within apocentre, though such constraints require observations over multiple orbital periods. We interpreted these results from the theoretical perspective of secular versus non-secular orbital dynamics, where a capturing of the former is required for good constraints on the overall amount of mass within apocentre, but in addition a capturing of the latter is required to yield unambiguous constraints on the radial profile.

\subsection{Outlook}

Concerning the prospects to improve the measurement uncertainties it is important to note that the uncertainty in RV is more constrained by the source S2 (that is, by the width of the measured spectral line) than by the instrument, such that ERIS does not bring a significant improvement over SINFONI \citep{Eisenhauer+2003_SINFONI, Bonnet+2004_SINFONI}. While the GRAVITY+ upgrade \citep{Eisenhauer2019_GRAVITY+} will drastically improve the sensitivity of the instrument with respect to fainter objects, a significant reduction of the astrometric uncertainties for S2 is not expected. In summary, the $50\mu''$ and $10\,\mathrm{km/s}$ uncertainties we assumed will remain realistic also in the upcoming years.

Since our above discussion was limited to the use of data for a single star (S2), it is reasonable to ask if the use of data for more stars would yield different conclusions. This is of interest in particular regarding the development of the Extremely Large Telescope (ELT). With its Multi-AO Imaging Camera for Deep Observations (MICADO) \citep{Davies+2018_MICADO} it is expected to track the S-stars with similar uncertainties to GRAVITY \citep{Pott+2018_MICADO}. Its much wider field of view will, however, allow it to observe a whole family of S-stars in a single frame, thus enabling it to gather data for more stars at a much higher rate than is currently possible. Nevertheless, assuming spherical symmetry, we do not expect to arrive at different conclusions than in our present discussion when incorporating data for more stars. 
% @Gernot: the following statements seem strong (disregard if you're sure)
This is because our results indicate that the limiting factor to capture the non-secular effects, and thus the radial profiles, is the measurement accuracy which data from more stars does not improve. For the constraints on the secular effect and thus on the overall amount of distributed mass within apocentre, we showed that the limiting factor is the number of observed periods, and hence also in this respect data for more stars is not helpful. It is true though, that different stellar orbits probe different radial domains, and therefore constrain the distribution slightly better in different regions. Also more eccentric orbits are more susceptible to the non-secular component of the dynamics. However the true benefit of multiple stellar data would only unfold if one drops the assumption of spherical symmetry, since then different orbital orientations become important.

In addition to improving the unbiased constraints on the overall amount of distributed matter within apocentre, future observations of the S-stars will certainly continue to improve biased constraints on specific classes of Dark Matter distributions -- a topic which has seen a strong very recent rise in interest \citep{FujitaCardoso2017, Ferreira+2017, Lacroix2018, Bar+2019_ultra_light_DM, GRAVITY+19_Scalar_fields, GRAVITY+2023_scalar_clouds, GRAVITY+22_mass_distribution, Nampalliwar21, Heissel+22, Chan2022, Yuan+2022_ultra_light_bosons, Bambhaniya+2022, DellaMonicaDeMartino2022, DellaMonicaDeMartino2023_ultra_light_bosons_1, DellaMonicaDeMartino2023_ultra_light_bosons_2, ChanLee2023, Shen+2023_DM_spikes}. Our results however underline the problem which this approach faces: its results all come with their own and distinct a~priori assumptions, and thus fail to be mutually exclusive. In other words, the biased approach does not contribute to a downsizing of the zoo of candidate models, but rather helps to restrict the parameter spaces of single candidate models. Despite being important, this task is of limited utility for answering the bigger question about the underlying nature of Dark Matter.

%\begin{acknowledgements}
%      .
%\end{acknowledgements}


\bibliography{BibFile}
\bibliographystyle{aa}

\end{document}